\documentclass[11pt,letterpaper]{article}
\usepackage[margin=1in]{geometry}
\usepackage[T1]{fontenc}
\usepackage{lmodern}
\usepackage[utf8]{inputenc}
\usepackage{amsmath,amssymb,amsthm}
\usepackage{mathtools,booktabs,graphicx,xcolor,enumitem}
\usepackage{flafter}
\usepackage[numbers,sort&compress]{natbib}
\AddToHook{env/thebibliography/begin}{\raggedright}
\usepackage{microtype}
\usepackage[colorlinks=true,linkcolor=blue!55!black,citecolor=blue!70!black,urlcolor=blue!60!black]{hyperref}
\usepackage{tikz}
\usetikzlibrary{arrows.meta,positioning}
\setlist{topsep=4pt,itemsep=2pt,parsep=0pt}
\newcommand{\OPT}{\operatorname{OPT}}
\newcommand{\supp}{\operatorname{supp}}
\newcommand{\Vcoef}{\mathcal V}
\newcommand{\R}{\mathbb R}
\newcommand{\E}{\mathbb E}
\newcommand{\ind}{\mathbf 1}

\newcommand{\pos}[1]{[#1]_+}

\newcommand{\eps}{\varepsilon}

\newcounter{algorithm}
\newcommand{\algorithmname}{Algorithm}

\newenvironment{manuscriptalgorithm}[1]{%
  \begin{figure}[tbp]\refstepcounter{algorithm}
  \hrule\vspace{4pt}
  \noindent\textbf{\algorithmname~\thealgorithm: #1}\par
  \vspace{4pt}\hrule\vspace{5pt}
  \begin{minipage}{\linewidth}\small
}{%
  \end{minipage}\par\vspace{5pt}\hrule
  \end{figure}
}
\newtheorem{theorem}{Theorem}[section]
\newtheorem{lemma}[theorem]{Lemma}
\newtheorem{proposition}[theorem]{Proposition}

\theoremstyle{definition}

\numberwithin{equation}{section}
\hypersetup{pdftitle={An Analytic 0.48-Approximation for Non-Monotone k-Submodular Maximization under a Matroid Constraint},pdfauthor={Zizong Wang, Yibo Wang, and Wenguo Yang}}
\title{An Analytic $0.48$-Approximation for Non-Monotone
$k$-Submodular Maximization under a Matroid Constraint}
\author{Zizong Wang \qquad Yibo Wang \qquad Wenguo Yang\\[5pt]
\small University of Chinese Academy of Sciences}
\date{}
\begin{document}
\maketitle
\begin{abstract}
We give a randomized polynomial-time $0.48$-approximation for maximizing a nonnegative non-monotone $k$-submodular function under a matroid constraint, for every $k\ge2$. The algorithm combines a top-two label rule with a nested-chain potential and maximum-weight matroid bases. Its analysis reduces the approximation guarantee to a one-dimensional time allocation: a uniform local lower bound leads to a linear differential equation, concavity closes the remaining interval, and two elementary budgets complete the proof. A strict saving in the time budget absorbs sampling and discretization errors, yielding the exact $0.48$ guarantee in the value-oracle model.
\end{abstract}
\vspace{0.6em}

% ---- sections/introduction-en.tex ----
\section{Introduction}\label{sec:introduction}

Submodularity expresses diminishing returns: the benefit of adding an element decreases as the chosen set grows. It supports a broad theory of approximation algorithms, from classical greedy methods to continuous optimization and rounding. A $k$-submodular function extends this model by assigning each selected element one of $k$ labels. A feasible solution is therefore a collection of $k$ disjoint sets, and a support constraint controls which elements may be selected, irrespective of their labels. Such models arise when a selection problem also includes an allocation or classification decision.

The label structure changes non-monotone optimization in a useful way. Although an individual label may have a negative marginal gain, the sum of the marginal gains of any two distinct labels is nonnegative. This pairwise monotonicity complements diminishing returns within an orthant. Together, the two properties characterize $k$-submodularity \cite{WZ2016}. They make it possible to compare an algorithmic label with an unknown optimal label while retaining information that would be lost by treating deletion and relabeling separately.

For unconstrained maximization, randomized label choices already yield a $1/2$-approximation \cite{ITY2016}, and better guarantees are known for $k\ge3$ \cite{Oshima2021}. Under a matroid constraint, however, the selection and labeling decisions must be coordinated. Greedy and continuous approaches long gave $1/3$-type guarantees \cite{SunLiuLi2022,NiuEtAl2023,HWZ2023}. Chen et al. recently broke this barrier with a proportional top-two rule, obtaining $\sqrt2-1$ \cite{ChenEtAl2026}. Aggarwal subsequently gave geometry-dependent continuous guarantees whose general-matroid benchmarks are $0.4456\ldots$ for $k=2$ and $0.4541\ldots$ for $k\ge3$ \cite{Aggarwal2026}. Our result improves both to $0.48$ with a common analysis for all $k\ge2$.

\begin{table}[t]
\centering
\caption{Selected guarantees for non-monotone $k$-submodular maximization under a general single matroid, through 10 October 2026.}\label{tab:progress}
\small
\begin{tabular}{@{}p{2.85in}p{1.22in}p{1.25in}@{}}
\toprule
Work & $k=2$ & $k\ge3$\\
\midrule
Earlier constrained $k$-submodular algorithms \cite{SunLiuLi2022,NiuEtAl2023} & $1/3$ or $1/3-\eps$ & $1/3$ or $1/3-\eps$\\
Huang--Wang--Zhou \cite{HWZ2023} & $1/3-\eps$ & $1/3-\eps$\\
Chen et al. \cite{ChenEtAl2026} & $\sqrt2-1$ & $\sqrt2-1$\\
Aggarwal \cite{Aggarwal2026} & $0.445601\ldots-\eps$ & $0.454126\ldots-\eps$\\
\textbf{This work} & $\boldsymbol{0.48}$ & $\boldsymbol{0.48}$\\
\bottomrule
\end{tabular}
\end{table}

\subsection{Our contribution}

Our contribution is an analytic approximation certificate. We combine the top-two label comparison with a joint nested-chain potential and choose the continuation budget through a single linear differential equation. A simple lower bound separates the two local parameters; a concavity argument handles the small slopes; and two elementary integrals certify that the construction fits into less than one unit of time. The strict time saving also supplies the margin needed for a finite value-oracle algorithm. Table~\ref{tab:progress} places the resulting guarantee in the current progression.

\begin{theorem}[Main result, informal]\label{thm:informal}
For every $k\ge2$, nonnegative non-monotone $k$-submodular maximization under a single matroid constraint admits a randomized polynomial-time $0.48$-approximation. The algorithm uses value and matroid-independence oracles.
\end{theorem}

\subsection{Technical overview}\label{sec:overview}

\paragraph{Keep the losses together.}
A standard proof gradually replaces an optimal solution by the algorithm's solution and charges the decrease in the comparison solution to the algorithm's gain. In the non-monotone setting, selecting an element may decrease this comparison even before any optimal element is removed. Charging these effects independently wastes the interaction between their labels. We instead average comparison solutions along one nested chain: as a common threshold increases, residual optimal elements disappear in a fixed order. Their marginal gains and the gains of the two algorithmic labels can then be compared on the same chain. One local inequality pays for relabeling, addition, and removal together.

\paragraph{Let the matroid do one job.}
The local comparison gives a nonnegative weight for each element. At each step, the algorithm selects a maximum-weight matroid basis and distributes a small amount of probability between the element's two best labels. The active residual part of the optimal solution is independent, so its total weight is at most the basis weight. This is the entire matroid comparison. The unknown optimal solution appears only in the proof; the algorithm uses the current gradient and the matroid oracle.

\paragraph{Turn continuation into a time budget.}
A scalar profile records how much comparison value can remain at a given selection probability. Its slope determines the label rule. A stage with parameter $b$ contributes a local coefficient to every smaller slope, so the desired approximation becomes a family of one-dimensional budget inequalities. Rather than fit many stage constants, we use a uniform lower bound for the local coefficient. On the high region, making the budget tight leads to a linear differential equation. On the low region, the coefficient is affine and the remaining budget is concave, so two endpoint checks suffice. Those checks reduce to a time integral and a value integral. Both have elementary proofs.

\paragraph{Finish with a finite algorithm.}
The continuous certificate leaves a small positive amount of time unused. A short additional stage turns it into a strict value margin. The profile can then be discretized and smoothed, and its gradients estimated by bounded marginal samples. The multilinear-extension machinery and value-preserving matroid rounding follow Huang, Wang, and Zhou \cite{HWZ2023}; Appendix~\ref{app:implementation} proves the additional estimates for our nested potential. This is why the final theorem has the exact factor $0.48$.

\subsection{Related work}\label{sec:related}

\paragraph{$k$-submodularity and label choices.}
Huber and Kolmogorov introduced the general $k$-submodular framework \cite{HK2012}; earlier structural work on bisubmodularity includes \cite{AFN1996}. Ward and \v Zivn\'y characterized the class by orthant submodularity and pairwise monotonicity and studied non-monotone maximization \cite{WZ2016}. Iwata, Tanigawa, and Yoshida gave a $1/2$-approximation in the unconstrained non-monotone case and a $k/(2k-1)$-approximation in the monotone case, together with oracle hardness \cite{ITY2016}. Oshima improved the unconstrained non-monotone ratio for $k\ge3$ \cite{Oshima2021}. Our use of two labels builds on this line of local comparisons and on the proportional rule of Chen et al. \cite{ChenEtAl2026}; the probabilities here are adapted to the continuation profile.

\paragraph{Support constraints.}
For monotone objectives, Ohsaka and Yoshida studied size constraints \cite{OhsakaYoshida2015}, and Sakaue gave a $1/2$-approximation under a matroid constraint \cite{Sakaue2017}. Combinatorial matroid algorithms were developed by Sun, Liu, and Li \cite{SunLiuLi2022}, and accelerated through decreasing thresholds by Niu et al. \cite{NiuEtAl2023}. Budgeted algorithms include Wang's $1/2$ monotone and $1/3$ non-monotone guarantees \cite{Wang2025}, the knapsack algorithms of Xiao et al. \cite{XiaoEtAl2026}, and simultaneous matroid and knapsack constraints \cite{YuEtAl2023}. Nguyen and Thai studied streaming maximization with noisy queries and size constraints \cite{NguyenThai2020}. Huang, Wang, and Zhou introduced a multilinear extension with gradient estimation and rounding for these problems \cite{HWZ2023}; its published development appears as \cite{ZHW2025}. Chen et al. obtained $\sqrt2-1$ for a single matroid and knapsack \cite{ChenEtAl2026}. Aggarwal's recent geometry-dependent analysis improves those guarantees and studies broader support regions \cite{Aggarwal2026}. We focus on an arbitrary single matroid and exploit its maximum-weight basis comparison.

\paragraph{Ordinary non-monotone submodular maximization.}
The continuous approach originates in the multilinear-extension and continuous-greedy work of Calinescu et al. \cite{CCPV2011}. Earlier constrained non-monotone algorithms include local search \cite{LeeMNS2009}, offline and secretary methods \cite{GRT2010}, and measured continuous greedy \cite{FNS2011}. Randomized double greedy gives the optimal $1/2$ ratio without constraints \cite{BFNS2015}. For a general matroid, later algorithms reached $0.372$ \cite{EneNguyen2016}, $0.385$ \cite{BuchbinderFeldman2019} and then $0.401$ \cite{BuchbinderFeldman2024}. The latter is the polynomial-time general-matroid benchmark for $k=1$ cited in \cite{FeldmanKuhnle2026}.

Recent research also improves determinism and oracle efficiency. Buchbinder and Feldman developed an extended multilinear representation and faster matroid algorithms \cite{BuchbinderFeldman2025}; Chen et al. gave a deterministic $0.385$-type matroid guarantee \cite{ChenGLSZ2026}. Guided combinatorial methods and reduced-query cardinality algorithms provide complementary tradeoffs \cite{ChenNPK2024,TukanMualemFeldman2024}; low-adaptivity algorithms offer another approach \cite{ChenChenKuhnle2025}. Poisson-process algorithms maintain an integral feasible set throughout optimization: Ganz Rozenman et al. treat monotone objectives \cite{GRKS2026}, and Kulik et al. extend this approach to a $1/e$ guarantee for non-monotone objectives \cite{KOSS2026}. Nematollahi, Vladu, and Zhao obtain $1-1/e-\eps$ with fixed-parameter dependence on matroid rank \cite{NematollahiVladuZhao2026}; this is a different running-time regime from polynomial dependence on rank. On the hardness side, the symmetry-gap framework \cite{Vondrak2013} yields a $0.478$ barrier through the construction of Oveis Gharan and Vondr\'ak \cite{GharanVondrak2011}, recently strengthened to $8/17$ even for cardinality and certain partition constraints \cite{FeldmanKuhnle2026}. These $k=1$ barriers coexist with our result because pairwise monotonicity is a structural property of the $k\ge2$ objective class. For $k=2$, the unconstrained $1/2$ oracle barrier \cite{WZ2016} also bounds what can hold for all matroids.

\paragraph{Rounding and our analytic reduction.}
Pipage rounding \cite{AS2004} and swap rounding \cite{CVZ2010} connect fractional optimization to discrete matroid feasibility. The general relaxation and contention-resolution framework is developed in \cite{CVZ2014}. We use the $k$-submodular version proved in \cite{HWZ2023}. The new part of the argument is the combination of the nested-chain comparison with an analytic continuation certificate. It avoids an optimized parameter table: the local lower bound determines the differential equation, and the two budgets certify the approximation.

\paragraph{Organization.}
Section~\ref{sec:prelim} fixes the notation. Sections~\ref{sec:local}--\ref{sec:framework} establish the local comparison, nested potential, and finite-certificate framework. Section~\ref{sec:analysis} gives the analytic $0.48$ certificate. The exact budget calculations and finite implementation are collected in the appendices.

% ---- sections/framework-en.tex ----
\section{Preliminaries}\label{sec:prelim}

Let $V$ be a ground set of size $n$, let $k\ge2$, and let
$\mathcal M=(V,\mathcal I)$ be a matroid given by an independence oracle.
An assignment $\boldsymbol s\in\{0,1,\ldots,k\}^{V}$ assigns a label to each
element; label zero means that the element is unused. Write
$\supp(\boldsymbol s)=\{e:s_e\ne0\}$. The order
$\boldsymbol s\preceq\boldsymbol t$ preserves every nonzero label of
$\boldsymbol s$. The meet $\boldsymbol s\sqcap\boldsymbol t$ preserves common
labels, and the join $\boldsymbol s\sqcup\boldsymbol t$ preserves a nonzero
label if the other label is zero or agrees with it, and assigns zero to
conflicting nonzero labels. A function $f:\{0,\ldots,k\}^{V}\to\R_{\ge0}$
is $k$-submodular if
\[
 f(\boldsymbol s)+f(\boldsymbol t)
 \ge f(\boldsymbol s\sqcap\boldsymbol t)+f(\boldsymbol s\sqcup\boldsymbol t).
\]
We maximize $f$ over assignments with support in $\mathcal I$, and denote the
optimum by $\OPT$. Values of $f$ are available through a value oracle.

\begin{theorem}\label{thm:main}
For every nonnegative $k$-submodular function with $k\ge2$ and every matroid,
there is a randomized algorithm using polynomially many value and
independence queries that returns an assignment $\boldsymbol s$ satisfying
\[
 \supp(\boldsymbol s)\in\mathcal I,
 \qquad \E[f(\boldsymbol s)]\ge\frac{12}{25}\OPT.
\]
\end{theorem}

Delete matroid loops, which cannot occur in a feasible support. If the rank
is zero, the zero assignment is optimal. Otherwise define
\[
 M=\max\bigl(\{0\}\cup
 \{f(i\ind_e)-f(\boldsymbol0):e\in V,\ i\in[k]\}\bigr).
\]
Here $i\ind_e$ denotes the assignment that labels only $e$ by $i$.
For a partial assignment with $s_e=0$, write
$\Delta_{e,i}f(\boldsymbol s)=f(\boldsymbol s+i\ind_e)-f(\boldsymbol s)$.
$k$-submodularity implies orthant submodularity and pairwise monotonicity:
\[
 \Delta_{e,i}f(\boldsymbol s)\ge\Delta_{e,i}f(\boldsymbol t)
 \quad(\boldsymbol s\preceq\boldsymbol t, t_e=0),
 \qquad
 \Delta_{e,i}f(\boldsymbol s)+\Delta_{e,j}f(\boldsymbol s)\ge0
 \quad(i\ne j).
\]
These follow by applying the defining inequality to two assignments that
differ by one addition, and to two different labels of the same element,
respectively. Consequently every marginal lies in $[-M,M]$. If $M=0$,
all marginals vanish and the zero assignment is optimal. Henceforth $M>0$;
since singleton supports are feasible and $f(\boldsymbol0)\ge0$, $M\le\OPT$.
Pairwise monotonicity also implies that an optimal assignment can be extended
to a matroid base without decreasing its value. Fix one such optimum
$\boldsymbol o$, its base $O$, and its labels $\ell_e=o_e$ for $e\in O$.

\subsection{The multilinear extension}

For a fractional vector $x=(x_{e,i})$, put $q_e=\sum_i x_{e,i}$ and
\[
 D=\{x\in\R_{\ge0}^{V\times[k]}:q_e\le1\text{ for every }e\}.
\]
Independently for each element, select label $i$ with probability $x_{e,i}$
and zero with probability $1-q_e$. Let $R_x$ be the resulting assignment and
define $F(x)=\E[f(R_x)]$. This is the multilinear extension of
\citet[Definition 3.2]{HWZ2023}.

\begin{lemma}[Multilinear properties; {\cite[Lemma 3.3]{HWZ2023}}]\label{lem:extension}
The function $F$ is affine in each row. On $D$,
\[
 |\partial_{e,i}F|\le M,\qquad
 \partial_{e,i}F+\partial_{e,j}F\ge0\quad(i\ne j),
 \qquad
 -2M\le\partial_{e,i}\partial_{a,j}F\le0\quad(e\ne a),
\]
and all second derivatives within one row vanish. In particular, $x\le y$
coordinatewise implies $\partial_{e,i}F(x)\ge\partial_{e,i}F(y)$.
Moreover, if $x,x+z\in D$ and $z\in\delta D$, then
\[
 F(x+z)\ge F(x)+\langle\nabla F(x),z\rangle-Mn^2\delta^2.
\]
\end{lemma}
\begin{proof}
The derivative $\partial_{e,i}F(x)$ is the expected marginal
$\Delta_{e,i}f(R_{x,-e})$, where row $e$ is omitted. Thus the marginal bounds
and pairwise monotonicity pass to $F$. A mixed derivative of two different
rows is the expected difference of two marginals. Orthant submodularity
makes it nonpositive, and the marginal bounds give its lower bound.
The expectation is affine in every row. Taylor's integral remainder and
$\|z\|_1\le n\delta$ give the finite-step inequality.
These are also the properties established in
\citet[Lemma 3.3 and Appendix B]{HWZ2023}.
\end{proof}

% ---- oracle-tools-en.tex ----
The following bounded-marginal version uses the sampling method of
\citet[Lemma 3.4 and Appendix A]{HWZ2023}.

\begin{lemma}[Bounded-marginal sampling]\label{lem:sampling-oracle}
Fix $x\in D$, $\alpha>0$, and $\delta\in(0,1)$. For each $(e,i)$, average
$J$ independent samples
$f(R_{x,-e}+i\ind_e)-f(R_{x,-e})$ to obtain $\widehat g_{e,i}$, where
\[
 J\ge\left\lceil\frac{2M^2}{\alpha^2}
                    \log\frac{2nk}{\delta}\right\rceil.
\]
Then
$\Pr\{\|\widehat g-\nabla F(x)\|_\infty\le\alpha\}\ge1-\delta$.
The estimates require at most $2nkJ$ value queries.
\end{lemma}
\begin{proof}
Each sample has mean $\partial_{e,i}F(x)$ and lies in $[-M,M]$.
Hoeffding's inequality bounds each coordinate's failure probability by
$2\exp(-J\alpha^2/(2M^2))\le\delta/(nk)$; a union bound proves the claim.
\end{proof}

\begin{lemma}[Single-matroid rounding,
{\citealp[Lemma 3.5 and Appendix C]{HWZ2023}}]\label{lem:matroid-rounding}
Let $f$ be a nonnegative, possibly non-monotone $k$-submodular function.
If $x\in D$ and its row-mass vector $q$ belongs to the matroid independent-set
polytope $P(\mathcal M)$, then a randomized polynomial-time algorithm
using independence queries and no value queries returns an assignment
$\boldsymbol s$ such that
\[
 \supp(\boldsymbol s)\in\mathcal I,
 \qquad \E[f(\boldsymbol s)]\ge F(x).
\]
\end{lemma}

\section{The joint local inequality}\label{sec:local}

For $0<b\le1/2$ and $0\le d\le b$, define
\begin{equation}\label{eq:kernel}
 \Vcoef(b,d)=
 \begin{cases}
 \displaystyle\frac{(1-b)(1-d)}{1+b},
 &d\le\displaystyle\frac{1-b}{3+b},\\[5pt]
 \displaystyle\frac2b\left(\sqrt{(1-b)d(b+d)}-d\right),
 &d\ge\displaystyle\frac{1-b}{3+b}.
 \end{cases}
\end{equation}
Only the part of either branch in $[0,b]$ is used. Given a vector $w$ of
upper bounds on the label marginals, relabel its largest two entries as
$w_1\ge w_2$. When both are positive, set $z=w_2/w_1$ and use the two labels
with probabilities $p$ and $1-p$, where
\begin{equation}\label{eq:top2}
 p=
 \begin{cases}
 1,&z\le(1-b)/(1+b),\\[2pt]
 \displaystyle\frac12+\frac{d}{2b}\frac{1-z}{1+z},
 &z>(1-b)/(1+b).
 \end{cases}
\end{equation}
For this rule define
\[
 \begin{aligned}
 G&=pw_1+(1-p)w_2, & U&=(2p-1)w_2,\\
 K&=\max\{(d+b-2bp)w_1,(d+2bp-b)w_2\}, & W&=(1-b)G-bU.
 \end{aligned}
\]
If $w_1>0\ge w_2$, use the largest label deterministically and set
$G=w_1$, $U=K=0$, $W=(1-b)w_1$. If $w_1\le0$, use a largest label and set
all four quantities to zero.

\begin{lemma}\label{lem:local}
The rule has $1/2\le p\le1$. Its quantities satisfy
\[
 (1-b)G-K\ge\Vcoef(b,d)\pos{w_1},\qquad
 K\ge bU\ge0,\qquad W\ge\Vcoef(b,d)\pos{w_1}\ge0.
\]
For fixed $b$, the function $\Vcoef(b,\cdot)$ is concave and decreasing on
$[0,b]$, and takes values in $[0,1]$.
\end{lemma}
\begin{proof}
Consider $w_1,w_2>0$ and normalize $w_1=1$. Put $h=(1-b)/(1+b)$.
For $z\le h$, $p=1$ and
$(1-b)G-K=(1-b)-(b+d)z\ge(1-b)(1-d)/(1+b)$.
For $z>h$, the two entries defining $K$ are equal. Substituting the rule
and writing $v=1+z$ gives
\[
 (1-b)G-K=\frac{(1-b)(b+d)}{2b}v+\frac{2d}{bv}-\frac{2d}{b},
 \qquad \frac2{1+b}\le v\le2.
\]
If $d>0$, its unconstrained minimum occurs at
$v_*=2\sqrt{d/((1-b)(b+d))}$. The assumption $d\le b\le1-b$ gives
$v_*\le2$, and $v_*\le2/(1+b)$ is equivalent to
$d\le(1-b)/(3+b)$. The constrained minimum is therefore precisely
\eqref{eq:kernel}. The case $d=0$ follows directly. The second entry of
$K$ equals $dw_2+bU$, so $K\ge bU$.

On the first branch, $\Vcoef_d=-(1-b)/(1+b)$. On the second, with
$Z=(1-b)d(b+d)$,
\[
 \Vcoef_d=\frac2b\left(\frac{(1-b)(2d+b)}{2\sqrt Z}-1\right),
 \qquad \Vcoef_{dd}=-\frac{(1-b)^2b}{2Z^{3/2}}\le0.
\]
The branches and their first derivatives agree at the boundary. They are
nonnegative because $Z-d^2=bd(1-b-d)\ge0$. Concavity makes the derivative
nonpositive and gives $\Vcoef(b,d)\le\Vcoef(b,0)=(1-b)/(1+b)\le1$.
The remaining sign cases follow from the definitions and
$\Vcoef(b,d)\le1-b$.
\end{proof}

\begin{lemma}[Joint chain bound]\label{lem:chain-local}
Suppose $w_1\ge w_2>0$ and $1/2\le p\le1$. Let $A(u)\in\R^k$,
$0\le u\le b$, be coordinatewise nondecreasing, with
$A_i(u)\le w_i$ and $A_i(u)+A_j(u)\ge0$ for distinct labels. Let
$D\in\R^k$ obey the same bounds and pairwise inequalities, and suppose
$A(u)\le D$ for $u<r$ and $A(u)\ge D$ for $u>r$, where $r\in[0,b]$.
For any label $\ell$ and $0\le d\le b$,
\[
 \int_0^r A_\ell(u)\,du+(d-r)D_\ell
 -\int_0^b\bigl(pA_1(u)+(1-p)A_2(u)\bigr)\,du\le K,
\]
where $K$ is as above. Moreover, the last integral is at least $-bU$.
If $w_1>0\ge w_2$ and $p=1$, both statements hold with $K=U=0$.
\end{lemma}
\begin{proof}
Put $s=2p-1$. For $\ell=1$, pairwise monotonicity and the crossing condition
give $A_2(u)\ge-D_1$ on both sides of $r$. Bounding $A_1$ by $D_1$ with
the appropriate signs on each interval gives an upper bound
$(d-bs)D_1$. If this coefficient is nonnegative, use $D_1\le w_1$;
otherwise use $D_1\ge-D_2\ge-w_2$. In either case the result is at most
$K$. For $\ell=2$, split the expression at $r$ as
\[
 p\int_0^r A_2-(1-p)\int_r^b A_2
 -p\int_0^b A_1+(d-r)D_2.
\]
The crossing condition bounds the first two integrals by
$(r-(1-p)b)D_2$, while pairwise monotonicity gives
$A_1(u)\ge-D_2$ on both intervals. The resulting bound is
$(d+bs)D_2\le(d+bs)w_2$.

For another label, write $D=D_\ell$. The comparison terms are at most $dD$,
and the same two-interval argument gives $A_1,A_2\ge-D$. If $D\ge0$,
then
$pA_1+(1-p)A_2=(1-p)(A_1+A_2)+sA_1\ge-sD$;
the result is at most $(d+bs)D\le(d+bs)w_2$. If $D<0$, the result is at
most $(d+b)D<0$.
Finally, $A_1\ge-A_2$ and $A_2\le w_2$ give the pure-addition bound
$pA_1+(1-p)A_2\ge-sw_2$.

When $w_2\le0$, every other label has nonpositive $A_i,D_i$, while
pairwise monotonicity forces $A_1,D_1\ge0$. The first-label calculation
becomes $(d-b)D_1\le0$, and the other-label calculation becomes
$(d+b)D_\ell\le0$. Pure addition is nonnegative.
\end{proof}

\section{A nested-chain potential}\label{sec:potential}

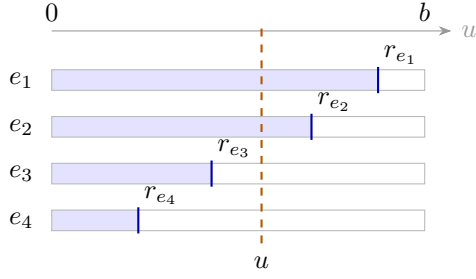
\begin{figure}[t]
\centering
\begin{tikzpicture}[x=0.88cm,y=0.62cm,>=Stealth,font=\small]
 \draw[->,gray!80] (0,1.05)--(6.0,1.05) node[right] {$u$};
 \node[above] at (0,1.05) {$0$};
 \node[above] at (5.6,1.05) {$b$};
 \foreach \y/\r/\j in {0/4.9/1,-1/3.9/2,-2/2.4/3,-3/1.3/4}{
  \fill[blue!11] (0,\y-0.22) rectangle (\r,\y+0.22);
  \draw[gray!55] (0,\y-0.22) rectangle (5.6,\y+0.22);
  \draw[blue!65!black,thick] (\r,\y-0.28)--(\r,\y+0.28);
  \node[left] at (-0.12,\y) {$e_\j$};
  \node[above right,inner sep=2pt] at (\r,\y+0.22) {$r_{e_\j}$};
 }
 \draw[dashed,thick,orange!70!black] (3.15,1.1)--(3.15,-3.55);
 \node[below] at (3.15,-3.55) {$u$};
\end{tikzpicture}
\caption{One threshold generates the nested comparison chain. Blue intervals indicate when an optimal row is filled, namely $u<r_e$. At the dashed threshold, only the first two rows are filled. Algorithmic and optimal label marginals are compared along this same chain.}

\label{fig:chain}
\end{figure}

For $x\in D$ and thresholds $r\in[0,1]^O$, fill the unused mass of selected
optimal rows by setting
\[
 h^u_{e,i}(x,r)=x_{e,i}+(1-q_e)\ind\{e\in O, i=\ell_e, u<r_e\},
 \qquad \Psi(x,r)=\int_0^1 F(h^u(x,r))\,du.
\]
Each filled vector belongs to $D$. The filled sets decrease as $u$
increases, so they form one nested chain; Figure~\ref{fig:chain} illustrates the shared threshold.

% ---- chain-figure.tex ----

\begin{lemma}[A support vector for the chain]\label{lem:support}
Order $O$ by nonincreasing thresholds, with arbitrary tie breaking, and let
$A_j$ be the first $j$ elements. Write $x_A$ for $x$ with rows in $A$
filled by their optimal labels. For the $j$th element $e$, put
\[
 D_{e,i}=\partial_{e,i}F(x_{A_{j-1}}),
 \qquad \xi_e=(1-q_e)D_{e,\ell_e}.
\]
For every $r'\in[0,1]^O$,
\[
 \Psi(x,r')\ge\Psi(x,r)+\sum_{e\in O}\xi_e(r'_e-r_e).
\]
Furthermore $A_{e,i}(u)=\partial_{e,i}F(h^u(x,r))$ and $D_{e,i}$ obey the
crossing and monotonicity conditions of Lemma~\ref{lem:chain-local}, with
crossing point $r_e$.
\end{lemma}
\begin{proof}
This is the chain support property of the Lov\'asz extension \cite{Lovasz1983}. For completeness, the set function $g(A)=F(x_A)-F(x)$ is submodular, by the nonpositive
mixed derivatives of different rows. Its chain increments are $\xi_e$.
For any set $S\subseteq O$, insert its elements in the chosen order;
submodularity gives $g(S)\ge\sum_{e\in S}\xi_e$. Integrating this inequality
for $S=\{e:r'_e>u\}$ gives
$\Psi(x,r')-F(x)\ge\sum_e r'_e\xi_e$.
For the original thresholds, each interval of positive length corresponds
to a prefix, where equality holds. This proves the support inequality.

As $u$ increases, $h^u$ decreases coordinatewise, so its gradient increases.
If $u<r_e$, the filled set contains every predecessor of $e$; if $u>r_e$,
it is contained in that predecessor set. Row $e$ itself does not affect its
own derivative. Lemma~\ref{lem:extension} gives the two crossing inequalities.
This reasoning applies to every ordering of tied thresholds.
\end{proof}

\section{From a profile to an algorithm}\label{sec:framework}

Within a stage with parameter $b\in(0,1/2]$, use a $C^{1,1}$ profile
$c(t,q)$ on $0\le q<1$ satisfying
\begin{equation}\label{eq:profile}
 0\le c\le b(1-q),\qquad
 0\le d:=-c_q\le b,\qquad
 -\Vcoef(b,d)\le c_t\le0.
\end{equation}
At points where $c=0$, require $c_t=c_q=0$. At state $x$, set
$r_e=c(t,q_e)/(1-q_e)$ for $e\in O$ and $d_e=-c_q(t,q_e)$ for all elements.
The algorithm uses $d_e$ and the current gradient estimates; it does not
compute $O$, the thresholds, or the potential.

\begin{lemma}[Weighted drift]\label{lem:drift}
Suppose the estimates obey
$\partial_{e,i}F(x)\le w_{e,i}\le\partial_{e,i}F(x)+2\eta$.
Compute each label distribution and weight by \eqref{eq:top2}, with parameters
$(b,d_e)$, and let $B$ be a maximum-weight matroid base. Put
$v_{e,i}=\ind\{e\in B\}p_{e,i}$ and
$\dot r_e=[c_t(t,q_e)+(r_e-d_e)\ind\{e\in B\}]/(1-q_e)$.
For any chain support vector of Lemma~\ref{lem:support},
\begin{equation}\label{eq:drift}
 D_x\Psi(x,r)[v]+\sum_{e\in O}\xi_e\dot r_e\ge-2\eta|B|.
\end{equation}
\end{lemma}
\begin{proof}
Let $R=\{e\in O:c(t,q_e)>0\}$. Rows outside $R$ have zero threshold and
zero threshold derivative. All true chain gradients are dominated by $w$,
since filling only increases coordinates, and they obey pairwise
monotonicity. The upper estimates themselves need no pairwise property.
For $e\in B\cap R$, differentiation with thresholds fixed contributes
\[
 \int_0^1\langle p_e,A_e(u)\rangle\,du
 -\int_0^{r_e}A_{e,\ell_e}(u)\,du.
\]
Adding the row-mass part $(r_e-d_e)D_{e,\ell_e}$ of the threshold term and
applying Lemma~\ref{lem:chain-local} bounds this below by
$(1-b)G_e-K_e-2\eta$. Here $h^u=x$ for $u>b$, which accounts for the
$G_e$ term. For $e\in B\setminus R$, the pure-addition part of that lemma
gives $W_e-2\eta$. If the maximum upper estimate is zero, pairwise
monotonicity forces all true current and filled gradients to vanish, so the
same conclusions hold.

Since $R$ is independent and weights are nonnegative, maximum-weight-base
optimality \cite{Edmonds1971} gives $\sum_{e\in B}W_e\ge\sum_{e\in R}W_e$. Thus the total
left side of \eqref{eq:drift} is at least
\[
 \sum_{e\in R\setminus B}W_e+
 \sum_{e\in R\cap B}\bigl((1-b)G_e-K_e\bigr)
 +\sum_{e\in R}c_t(t,q_e)D_{e,\ell_e}-2\eta|B|.
\]
Each first or second summand is at least $\Vcoef(b,d_e)w_{e,1}$ by
Lemma~\ref{lem:local}. Moreover $w_{e,1}\ge0$, $D_{e,\ell_e}\le w_{e,1}$,
and $c_t\le0$. Therefore its sum with the time term is at least
$(\Vcoef(b,d_e)+c_t)w_{e,1}\ge0$.
\end{proof}

\subsection{Finite certificates}

Choose rational stage parameters $0=b_0<b_1<\cdots<b_L\le1/2$, rational
durations $\tau_l\ge0$, and rational coefficients
$0\le\nu_{lj}\le\Vcoef(b_l,b_j)$ for $0\le j<l\le L$. Their certificate
value is
\begin{equation}\label{eq:certificate}
 \Gamma=\min_{0\le j\le L}
 \left\{b_j+\sum_{l=j+1}^L\tau_l\nu_{lj}\right\},
 \qquad T=\sum_{l=1}^L\tau_l<1.
\end{equation}
Execute stages in the order $L,L-1,\ldots,1$, skipping zero durations. In
stage $l$, let $R_l(t)$ be its remaining time and define
\[
 \begin{aligned}
 P_{lj}(t,q)&=b_j(1-q)+R_l(t)\nu_{lj}
       +\sum_{h=j+1}^{l-1}\tau_h\nu_{hj}\qquad(j<l),\\
 P_{ll}(t,q)&=b_l(1-q),
 \qquad \bar c(t,q)=\min_{0\le j\le l}P_{lj}(t,q).
 \end{aligned}
\]
This profile is continuous across stage boundaries: the old planes with
$j<l$ become the next stage's planes, and at the boundary the plane with $j=l-1$ equals the new cap
$b_{l-1}(1-q)$, which is no greater than the deleted cap for $q\le1$. It satisfies $\bar c(0,0)\ge\Gamma$,
$0\le\bar c\le b_l(1-q)$, and $\bar c(T,q)=0$.

\begin{theorem}[Finite-certificate realization]\label{thm:certificate}
For every fixed certificate \eqref{eq:certificate} and every
$\eps>0$, there is a randomized algorithm using
$\operatorname{poly}(n,k,1/\eps)$ value and independence queries whose output
is feasible and satisfies
\[
 \E[f(\boldsymbol s)]\ge(\Gamma-\eps)\OPT.
\]
\end{theorem}

Algorithm~\ref{alg:certificate} gives the finite implementation. Its smoothing,
sampling, and step-size bounds are proved in Appendix~\ref{app:implementation}.

% ---- algorithm-en.tex ----
\begin{manuscriptalgorithm}{Finite-certificate algorithm}
\label{alg:certificate}
\small
\noindent\textbf{Input:} A value oracle for nonnegative $k$-submodular $f$;
a matroid independence oracle; $k\ge2$; $\eps>0$; a fixed rational
certificate $(b_l,\tau_l,\nu_{lj})$ satisfying \eqref{eq:certificate}.
\par\noindent\textbf{Output:} A feasible assignment $\boldsymbol s$ with
$\E[f(\boldsymbol s)]\ge(\Gamma-\eps)\OPT$.
\begin{enumerate}[label=\arabic*:,leftmargin=1.8em,itemsep=1pt,topsep=3pt]
\item Delete matroid loops and let $n$ be the remaining ground-set size.
If the rank is zero or $\Gamma\le\eps$, \textbf{return} $\boldsymbol0$.
\item Compute $M=\max\bigl(\{0\}\cup
\{f(i\ind_e)-f(\boldsymbol0):e\in V,i\in[k]\}\bigr)$.
If $M=0$, \textbf{return} $\boldsymbol0$.
\item Set $a_0=(1-T)/2$, and choose
\[
 \theta=\frac{\eps}{4(nL+1)},\quad
 \eta=\frac{\eps M}{8n},\quad
 K_0=\frac{64n^2(L+1)}{\theta a_0^2},\quad h=\frac{\eps}{4K_0}.
\]
\item Set $m_l=\lceil\tau_l/h\rceil$ for positive-duration stages
and $m_l=0$ otherwise; set $m=\sum_lm_l$ and
$J=\left\lceil(2M^2/\eta^2)\log(8nkm/\eps)\right\rceil$.
Initialize $x=0$, $t=0$, and an empty list of weighted bases.
\item \textbf{For} $l=L,L-1,\ldots,1$ with $\tau_l>0$ \textbf{do}
\item \hspace*{1em}Reset the stage profile to the planes $P_{lj}$ above;
set $\Delta=\tau_l/m_l$ and remaining time $R_l=\tau_l$.
Use the explicit smoothing of Appendix~\ref{app:implementation},
$c=\psi_\theta\!\left(\min_{\omega\in\Delta_{l+1}}
\{\sum_j\omega_jP_{lj}+\tfrac\theta2\|\omega\|_2^2\}-\tfrac\theta2\right)$.
\item \hspace*{1em}\textbf{Repeat} $m_l$ \textbf{times}
\item \hspace*{2em}For each $e$, compute $q_e=\sum_i x_{e,i}$ and
$d_e=-\partial_qc(t,q_e)$.
\item \hspace*{2em}Using Lemma~\ref{lem:sampling-oracle}, for each $(e,i)$ draw $J$ independent assignments
$R_{x,-e}^{(j)}$ of all other rows and set
\[
 w_{e,i}=\eta+\frac1J\sum_{j=1}^J
 \bigl[f(R_{x,-e}^{(j)}+i\ind_e)-f(R_{x,-e}^{(j)})\bigr].
\]
\item \hspace*{2em}Order each row's estimates and apply
\eqref{eq:top2} with $(b_l,d_e)$ to obtain $p_{e,i}$ (zero on other labels) and
$W_e=(1-b_l)G_e-b_lU_e$.
Use the largest label deterministically if $w_{e,2}\le0$;
set $W_e=0$ if $w_{e,1}\le0$, as in Section~\ref{sec:local}.
\item \hspace*{2em}Find a maximum-weight matroid base $B$ for weights
$W_e$ by matroid greedy; set $v_{e,i}=\ind\{e\in B\}p_{e,i}$.
\item \hspace*{2em}Update $x\gets x+\Delta v$, $t\gets t+\Delta$,
$R_l\gets R_l-\Delta$; retain the base $B$ with coefficient $\Delta$.
\item \hspace*{1em}\textbf{End repeat}
\item \textbf{End for}
\item Recompute $q_e=\sum_i x_{e,i}$. Independently choose label $I_e=i$
with probability $x_{e,i}/q_e$
when $q_e>0$; choose any label when $q_e=0$.
\item Add as many free dummy elements as the matroid rank and truncate to that rank.
Add the all-dummy base with coefficient $1-T$ to the retained decomposition;
apply randomized swap rounding (Lemma~\ref{lem:matroid-rounding} and
Appendix~\ref{app:implementation}) and delete dummy elements to obtain $S$.
\item \textbf{Return} $s_e=I_e$ for $e\in S$ and $s_e=0$ otherwise.
\end{enumerate}
\end{manuscriptalgorithm}

The finite-certificate theorem converts the analytic construction of the
next section into the algorithm of Theorem~\ref{thm:main}: use its fixed
certificate with $\Gamma>0.48$ and set $\eps=(\Gamma-0.48)/2$.

% ---- sections/analysis-en.tex ----
\section{An analytic certificate for \texorpdfstring{$0.48$}{0.48}}\label{sec:analysis}

The framework leaves one task: allocate less than one unit of time so that every profile slope receives a budget of at least $0.48$. We solve this task analytically. Throughout this section,
\[
  \rho=\frac{12}{25},\qquad a=\frac15.
\]
The role of $a$ is simply to separate a region where the local coefficient is affine from a region governed by a linear differential equation.

\subsection{A uniform local lower bound}

\begin{lemma}\label{lem:uniform}
For $0<d\le b\le 1/2$,
\begin{equation}\label{eq:uniform}
 \Vcoef(b,d)\ge \lambda(d)(1-b-d),
 \qquad \lambda(d)=\frac{4\sqrt d}{(1+\sqrt d)^2}.
\end{equation}
\end{lemma}
\begin{proof}
The second expression in \eqref{eq:kernel} is the unconstrained minimum in the local calculation, so it is a lower bound for either branch. Rationalizing it gives
\[
 \Vcoef(b,d)\ge
 \frac{2\bigl(\sqrt{(1-b)d(b+d)}-d\bigr)}b
 =\frac{2(1-b-d)}{1+\sqrt{(1-b)(b+d)/d}}.
\]
The concave quadratic $(1-b)(b+d)$ is at most $(1+d)^2/4$. Substitution proves the claim. At $b=d=1/2$, both sides vanish and the same identity applies.
\end{proof}

This bound separates the slope $d$ from the stage parameter $b$. That separation is what makes the time allocation one-dimensional.

\subsection{One linear differential equation}

Put
\[
 H(d)=\frac{\rho-d}{\lambda(d)}\quad (a\le d\le\rho).
\]
To make \eqref{eq:uniform} meet the target exactly, we seek a positive density $\tau$ satisfying
\begin{equation}\label{eq:integral-identity}
 \int_d^\rho \tau(b)(1-b-d)\,db=H(d).
\end{equation}
Differentiating twice turns this identity into the linear equation
\begin{equation}\label{eq:ode}
 -(1-2b)\tau'(b)+3\tau(b)=H''(b),
 \qquad
 \tau(\rho)=\frac{1}{\lambda(\rho)(1-2\rho)}.
\end{equation}
There is no parameter search. Its solution is
\begin{equation}\label{eq:density}
 \tau(b)=
 \begin{cases}
 \displaystyle\frac{1+b}{2(1-b)^3},&0\le b<a,\\[6pt]
 \displaystyle\frac{\sqrt{1-2\rho}/\lambda(\rho)
       +\int_b^\rho\sqrt{1-2u}\,H''(u)\,du}
       {(1-2b)^{3/2}},&a\le b\le\rho.
 \end{cases}
\end{equation}
The value assigned at the single breakpoint does not affect any integral. Positivity follows from
\[
 H''(d)=\frac{3\rho+(1-\rho)d-3d^2}{16d^{5/2}}>0
 \qquad(a\le d\le\rho).
\]

\begin{lemma}\label{lem:high}
The density in \eqref{eq:density} satisfies \eqref{eq:integral-identity}. Consequently,
\[
 d+\int_d^\rho \tau(b)\Vcoef(b,d)\,db\ge\rho
 \qquad(a\le d\le\rho).
\]
\end{lemma}
\begin{proof}
Let $I(d)$ denote the integral on the left of \eqref{eq:integral-identity}. Direct differentiation and \eqref{eq:ode} give
\[
 I''(d)=-(1-2d)\tau'(d)+3\tau(d)=H''(d).
\]
At the right endpoint, $I(\rho)=H(\rho)=0$ and
$I'(\rho)=-(1-2\rho)\tau(\rho)=-1/\lambda(\rho)=H'(\rho)$.
Thus $I=H$. Apply Lemma~\ref{lem:uniform} and multiply the identity by $\lambda(d)$.
\end{proof}

\subsection{Concavity closes the low region}

For $0\le d\le b\le a$, only the affine branch of \eqref{eq:kernel} occurs. Hence the first line of \eqref{eq:density} gives the exact calculation
\[
 \int_d^a\tau(b)\Vcoef(b,d)\,db
 =\frac{a-d}{2(1-a)}.
\]
It follows that, for $0\le d\le a$, the entire budget is
\[
 d+\int_d^\rho\tau(b)\Vcoef(b,d)\,db
 =\frac18+\frac{3d}{8}
   +\int_a^\rho\tau(b)\Vcoef(b,d)\,db.
\]
This is concave in $d$, by Lemma~\ref{lem:local}. Its value at $a$ is at least $\rho$ by Lemma~\ref{lem:high}. To close the other endpoint, it is enough to prove
\begin{equation}\label{eq:two-budgets}
 \int_a^\rho\tau(b)\,db<\frac{27}{32},
 \qquad
 \int_a^\rho\tau(b)\frac{1-b}{1+b}\,db>\frac{71}{200}.
\end{equation}
The first inequality is the time budget; the second is the value budget at $d=0$. These are the only two scalar checks needed.

\begin{lemma}[The two budgets]\label{lem:budgets}
Both inequalities in \eqref{eq:two-budgets} hold. In fact,
\[
 \frac{21}{25}<\int_a^\rho\tau(b)\,db<\frac{27}{32}.
\]
\end{lemma}
\begin{proof}
Write $m=\int_a^\rho\tau(b)\,db$. Integrating \eqref{eq:density} gives an expression in radicals and one arctangent. A half-angle substitution followed by the first three terms of the alternating arctangent series gives $21/25<m<27/32$. Appendix~\ref{app:budgets} records the evaluation and the rational checks.

For the value budget, the identity
\[
 \frac{1-b}{1+b}
 =\frac12-\frac98\left(b-\frac13\right)
   +\frac{2(b-1/3)^2}{(4/3)^2(1+b)}
\]
gives the quadratic lower bound
\[
 \frac{1-b}{1+b}\ge\frac12-\frac98\left(b-\frac13\right)
  +\frac{2(b-1/3)^2}{(4/3)^2(1+\rho)}
 \qquad(a\le b\le\rho).
\]
Only the first two moments of $\tau$ are therefore needed. From \eqref{eq:integral-identity}, evaluated at $a$ and then integrated over $d\in[a,\rho]$, they are
\[
 \int_a^\rho b\tau(b)\,db=\frac45m-H(a),
 \qquad
 \int_a^\rho b^2\tau(b)\,db
 =\frac{13}{25}m-\frac45H(a)-\frac23\int_a^\rho H(d)\,dd.
\]
Substitution into the quadratic, using $m>21/25$, gives the value budget. This final check needs only the elementary evaluations of $H(a)$ and $\int_a^\rho H(d)\,dd$, and two radical bounds. The exact arithmetic is in Appendix~\ref{app:budgets}.
\end{proof}

\begin{proposition}[Continuous certificate]\label{prop:continuous}
The density \eqref{eq:density} satisfies
\begin{equation}\label{eq:continuous-certificate}
 d+\int_d^\rho\tau(b)\Vcoef(b,d)\,db\ge\rho
 \quad(0\le d\le\rho),\qquad
 T:=\int_0^\rho\tau(b)\,db<1.
\end{equation}
\end{proposition}
\begin{proof}
The high region is Lemma~\ref{lem:high}. In the low region, concavity reduces the claim to the two endpoints. At zero, Lemma~\ref{lem:budgets} gives
$1/8+71/200=12/25$. Finally,
\[
 \int_0^a\tau(b)\,db=\frac{a}{2(1-a)^2}=\frac5{32},
\]
so the time budget in Lemma~\ref{lem:budgets} gives $T<1$.
\end{proof}

\subsection{Turning the strict time saving into an exact guarantee}

The strict time saving in \eqref{eq:continuous-certificate} has an algorithmic purpose: it pays for finite implementation. Prepend a stage with parameter $b=1/2$ and duration
\[
 \delta=\frac{1-T}{2}.
\]
The total duration becomes $(1+T)/2<1$. Since $\Vcoef(1/2,d)$ is decreasing in $d$, every slope $d\le\rho$ gains at least
\[
 \eta_*=\delta\Vcoef(1/2,\rho)>0.
\]
For $\rho\le d\le1/2$, the budget $d+\delta\Vcoef(1/2,d)$ is concave. Its endpoint values are $\rho+\eta_*$ and $1/2$, respectively. Both are at least $\rho+\eta_*$: indeed $\Vcoef(1/2,\rho)\le1/2$ and $\delta< (1-5/32-21/25)/2<1/100$, so $\eta_*<1/50$.
Thus the augmented certificate has a strict value margin over $\rho$ for every slope in $[0,1/2]$.

The slack can also be bounded by explicit rational constants.
Appendix~\ref{app:budgets} proves $m<1687/2000$, so
$T<3999/4000$. One may therefore replace the additional duration by
$\delta_0=1/8000$, obtaining total duration less than $7999/8000$.
Moreover,
\[
 \Vcoef(1/2,12/25)=\frac{28\sqrt3-48}{25}
 >\frac{12}{625},
\]
where the strict inequality follows from $\sqrt3>433/250$ by squaring.
The same endpoint argument then gives a uniform value margin greater
than $3/1250000$. Thus a rational discretization with total budget error
less than $3/2500000$ suffices to retain a strict margin. These bounds
quantify the existence argument; they are not a practical stage-count bound.

\begin{proof}[Proof of Theorem~\ref{thm:main}]
Keep the prepended stage separately, and discretize the density $\tau$ on a sufficiently fine rational grid in $[0,\rho]$, including $a$ and $\rho$. Include $1/2$ as the final stage parameter. The density is bounded on each of its two intervals, and $\Vcoef$ is continuous on $\{(b,d):0\le d\le b\le1/2\}$ after defining $\Vcoef(0,0)=1$. Uniform continuity controls cells away from the moving lower endpoint $b=d$; its boundary cell contributes at most the mesh width times $\sup\tau$. Thus the discrete budgets converge uniformly to the continuous ones. Approximate stage durations and local coefficients by nonnegative rationals from below. The resulting finite certificate still has total duration less than one and value $\Gamma>\rho$.

All constants in this certificate are independent of the input. Apply Theorem~\ref{thm:certificate} with a fixed accuracy smaller than $\Gamma-\rho$. The resulting algorithm uses polynomially many value and independence queries and returns a feasible assignment $\boldsymbol s$ with
\[
 \mathbb E[f(\boldsymbol s)]\ge\rho\,\OPT=0.48\,\OPT.
\]
\end{proof}

The analysis uses an explicit local coefficient, one linear differential equation, concavity, and two elementary budgets. The extra stage comes directly from the unused time; it introduces no fitted parameter.

% ---- sections/conclusion-en.tex ----
\section{Concluding remarks}

The approximation factor follows from a short analytic construction. The nested chain keeps the losses of changing labels and removing optimal elements in the same comparison, while the matroid enters through maximum-weight bases. Once these structural ingredients are in place, the remaining calculation is a linear differential equation followed by concavity and two scalar budgets.

There remains a gap between $0.48$ and the $1/2$ oracle barrier for bisubmodular maximization. Closing it may require a stronger local comparison, or a time allocation that uses more of the exact coefficient than the separated lower bound does. It would also be useful to find a combinatorial implementation with a comparable guarantee and a substantially smaller query complexity.

\section*{Use of AI-assisted tools}
ChatGPT was used during preparation of this version to assist with
LaTeX formatting, bibliography reconstruction from the supplied manuscript,
proof checking, and supplementary algebraic and numerical checks.
Responsibility for the mathematical claims and references rests with the authors.

\appendix

% ---- sections/budgets-en.tex ----
\section{Exact evaluation of the two budgets}\label{app:budgets}

This appendix records the elementary arithmetic behind Lemma~\ref{lem:budgets}. It is independent of the finite algorithm.

\subsection{The time budget}

Integrating the defining solution \eqref{eq:density} in the opposite order gives
\[
 m=\frac{1-\sqrt{(1-2\rho)/(1-2a)}}{\lambda(\rho)}
   +\int_a^\rho
    \left(1-\sqrt{\frac{1-2u}{1-2a}}\right)H''(u)\,du.
\]
Integration by parts and the substitution
$z=\sqrt{(1-2u)/u}$ give
\begin{equation}\label{eq:mass}
 m=\frac12+\frac{49\sqrt5}{400}-\frac{\sqrt{15}}{30}
   +\frac{127\sqrt{30}}{2400}\arctan\sqrt{\frac23}.
\end{equation}
These operations involve only powers and a rescaled arctangent.

Set $y=\sqrt2/(\sqrt3+\sqrt5)$. The half-angle identity gives
$\arctan\sqrt{2/3}=2\arctan y$, and $0<y<1$. Integrating
\[
 1-z^2<\frac{1}{1+z^2}<1-z^2+z^4
 \qquad(0<z<1)
\]
from zero to $y$ gives
\[
 y-\frac{y^3}{3}<\arctan y<y-\frac{y^3}{3}+\frac{y^5}{5}.
\]
After substitution into \eqref{eq:mass} and simplification,
\[
 \frac12+\frac{303\sqrt5}{400}-\frac{127\sqrt3}{180}
              -\frac{\sqrt{15}}{30}
 <m<
 \frac12+\frac{3683\sqrt3}{720}-\frac{3751\sqrt5}{1000}
              -\frac{\sqrt{15}}{30}.
\]
For the lower bound, use
$\sqrt5>559/250$, $\sqrt3<26/15$, and $\sqrt{15}<31/8$; their validity follows by squaring. The lower expression then exceeds
\[
 \frac{2272429}{2700000}>\frac{21}{25}.
\]
For the upper bound, use
$\sqrt3<362/209$, $\sqrt5>682/305$, and $\sqrt{15}>213/55$. Again, squaring verifies all three inequalities. The upper expression is less than
\[
 \frac{483852433}{573705000}<\frac{1687}{2000}<\frac{27}{32}.
\]
These checks establish the strict time budget using rational arithmetic.

\subsection{The value budget}

The quadratic lower bound in Lemma~\ref{lem:budgets} expands to $(284-483b+225b^2)/296$. Integrating it and substituting the two moment identities gives
\begin{equation}\label{eq:value-budget}
 \int_a^\rho\tau(b)\frac{1-b}{1+b}\,db
 \ge\frac{\frac{73}{5}m+303H(a)-150\int_a^\rho H(d)\,dd}{296}.
\end{equation}
The two elementary quantities here are
\[
 H(a)=\frac7{50}+\frac{21\sqrt5}{250},\qquad
 \int_a^\rho H(d)\,dd
  =\frac{49}{2500}+\frac{1096\sqrt3}{15625}
                     -\frac{82\sqrt5}{1875}.
\]
For example, the second identity follows by expanding
\[
 H(d)=\frac14\left(\rho d^{-1/2}+2\rho
                +(\rho-1)\sqrt d-2d-d^{3/2}\right)
\]
and integrating term by term. Using $m>21/25$, the right-hand side of \eqref{eq:value-budget} exceeds
\[
 \frac{\frac{6468}{125}+\frac{8003\sqrt5}{250}
                       -\frac{6576\sqrt3}{625}}{296}.
\]
The same bounds $\sqrt5>559/250$ and $\sqrt3<26/15$ show that this is greater than
\[
 \frac{6567837}{18500000}
 =\frac{71}{200}+\frac{337}{18500000}.
\]
This proves the strict value budget. No numerical quadrature or optimization is used in either check.

% ---- sections/implementation-en.tex ----
\section{Finite implementation}\label{app:implementation}

This appendix proves Theorem~\ref{thm:certificate}. Fix its rational
certificate and a rational $a_0\in(0,1)$ with $T\le1-a_0$. Constants below
may depend on this fixed certificate. During the algorithm each row mass
satisfies $q_e\le t\le T$, so every denominator $1-q_e$ is at least $a_0$.
If $\Gamma\le\eps$, return the zero assignment; its nonnegative value
already satisfies the conclusion. Thus the construction below can assume
$0<\eps<\Gamma\le b_L\le1/2$ and $T>0$.

\subsection{An explicit smooth profile}

For $0<\theta\le1$, in stage $l$ define
\[
 s(t,q)=\min_{\omega\in\Delta_{l+1}}
 \left\{\sum_{j=0}^l\omega_jP_{lj}(t,q)
       +\frac\theta2\sum_{j=0}^l\omega_j^2\right\}-\frac\theta2,
\]
where $\Delta_{l+1}$ is the probability simplex. Set
\[
 \psi_\theta(s)=
 \begin{cases}0,&s\le0,\\s^2/(2\theta),&0<s<\theta,\\
 s-\theta/2,&s\ge\theta,\end{cases}
 \qquad c(t,q)=\psi_\theta(s(t,q)).
\]
The unique minimizing weights are
$\omega_j=\pos{\lambda-P_{lj}}/\theta$, where
$\sum_j\pos{\lambda-P_{lj}}=\theta$.
Sorting the planes determines $\lambda$ by a linear equation. Thus this
smoothing is computable by finite arithmetic operations.

\begin{lemma}\label{lem:smoothing}
Within each stage the smooth profile satisfies \eqref{eq:profile}, is
$C^{1,1}$, and obeys
\[
 \bar c-\theta\le c\le\bar c,\qquad
 |c_t|\le1,\quad |c_q|\le1/2,\quad
 \operatorname{Lip}(\nabla c)\le\frac{4(L+1)}\theta.
\]
Its two one-sided values at a stage boundary differ by at most $\theta$.
Also $c(0,0)\ge\Gamma-\theta$ and $c(T,q)=0$.
\end{lemma}
\begin{proof}
The regularized minimum lies between $\bar c-\theta/2$ and $\bar c$.
Since $0\le\psi_\theta(s)\le\pos{s}$ and
$\psi_\theta(s)\ge s-\theta/2$, the approximation and cap bounds follow.
Writing $\alpha=\psi'_\theta(s)\in[0,1]$, differentiation of the unique
minimum gives
\[
 d=-c_q=\alpha\sum_{j=0}^l\omega_jb_j,\qquad
 -c_t=\alpha\sum_{j=0}^{l-1}\omega_j\nu_{lj}.
\]
Set $\nu_{ll}=0$. Concavity and nonnegativity of $\Vcoef(b_l,\cdot)$ give
\[
 -c_t\le\alpha\sum_j\omega_j\Vcoef(b_l,b_j)
 \le\Vcoef\left(b_l,\alpha\sum_j\omega_jb_j\right)=\Vcoef(b_l,d).
\]
The second inequality includes additional weight $1-\alpha$ at zero.
If $c=0$, then $s\le0$ and $\alpha=0$, so both derivatives vanish.

The weights are the Euclidean projection of $(-P_{lj}/\theta)_{j=0}^l$ onto the simplex;
this projection is nonexpansive. Each plane gradient has norm at most
$\sqrt5/2$, giving
$\operatorname{Lip}(\nabla s)\le5(L+1)/(4\theta)$ and
$\|\nabla s\|_2\le\sqrt5/2$. Since
$\operatorname{Lip}(\psi'_\theta)=1/\theta$, the stated bound for $c$
follows. At a boundary, both smoothed values belong to the same interval
$[\bar c-\theta,\bar c]$, proving the jump bound.
\end{proof}

\subsection{Potential estimates and a finite step}

\begin{lemma}\label{lem:potential-estimates}
For fixed thresholds $r$, $\Psi(\cdot,r)$ is smooth. For $x,x+z\in D$,
\[
 \Psi(x+z,r)\ge\Psi(x,r)+D_x\Psi(x,r)[z]-4M\|z\|_1^2.
\]
For the support vector determined by the same threshold ordering,
$|\xi_e(x+z)-\xi_e(x)|\le5M\|z\|_1$.
If $r_e=c_e/(1-q_e)$ and $r'_e=c'_e/(1-q_e)$, then
\[
 |\Psi(x,r')-\Psi(x,r)|\le M\sum_{e\in O}|c'_e-c_e|.
\]
\end{lemma}
\begin{proof}
For a fixed filled set, the affine map $x\mapsto x_A$ expands the
$\ell_1$ norm of a perturbation by at most two. Lemma~\ref{lem:extension}
bounds the absolute second directional derivative of $F$ by
$2M\|z\|_1^2$. Taylor's integral remainder, followed by integration over
the threshold intervals, proves the first assertion. The factor
$1-q_e$ in $\xi_e$ changes by at most $\|z\|_1$, and its gradient factor
changes by at most $2M\cdot2\|z\|_1$. This gives the second assertion.
Changing one threshold alters an integral over an interval of length
$|r'_e-r_e|$; filling that row changes $F$ by at most $(1-q_e)M$.
Changing thresholds one at a time proves the last assertion.
\end{proof}

Let
\[
 K_0=\frac{64n^2(L+1)}{\theta a_0^2}.
\]
The threshold function $r(t,q)=c(t,q)/(1-q)$ satisfies, within one stage,
\[
 |r_t|\le a_0^{-1},\qquad |r_q|\le(2a_0)^{-1},\qquad
 \operatorname{Lip}(\nabla r)\le\frac{16(L+1)}{\theta a_0^2}.
\]
Indeed $r_t=c_t/(1-q)$ and $r_q=(r-d)/(1-q)$, and $r,d\in[0,1/2]$.
Differentiating once more almost everywhere, using Lemma~\ref{lem:smoothing}
and $c\le(1-q)/2$, bounds the Hessian norm by
$4(L+1)/(\theta a_0)+4/a_0^2$.

\begin{lemma}[Finite-step bound]\label{lem:finite-step}
Assume the upper-gradient event in Lemma~\ref{lem:drift}. Take its direction
$v$, a step length $\Delta>0$ remaining within one stage, and
$x'=x+\Delta v$, $t'=t+\Delta$. If $r,r'$ are the respective thresholds,
then
\[
 \Psi(x',r')\ge\Psi(x,r)-2\eta|B|\Delta-K_0M\Delta^2.
\]
\end{lemma}
\begin{proof}
Use any ordering of the old thresholds. Lemma~\ref{lem:support}, at $x'$,
and Lemma~\ref{lem:potential-estimates} give
\[
 \Psi(x',r')\ge\Psi(x,r)+\Delta D_x\Psi(x,r)[v]
 +\sum_e\xi_e(x)(r'_e-r_e)
 -4M\|x'-x\|_1^2-5M\|x'-x\|_1\|r'-r\|_1.
\]
We have $\|x'-x\|_1=|B|\Delta\le n\Delta$ and
$\|r'-r\|_1\le2n\Delta/a_0$. The threshold Taylor estimate gives
\[
 |r'_e-r_e-\Delta\dot r_e|
 \le\frac{16(L+1)}{\theta a_0^2}\Delta^2.
\]
Since $|\xi_e|\le M$, the total second-order cost is at most
\[
 M\Delta^2\left(4n^2+\frac{10n^2}{a_0}
                +\frac{16n(L+1)}{\theta a_0^2}\right)
 \le K_0M\Delta^2.
\]
Lemma~\ref{lem:drift} bounds the remaining first-order term. The support
inequality handles tied thresholds directly; no differentiability of
$\Psi$ in $r$ is needed.
\end{proof}

\subsection{Sampling and the error budget}

Choose
\[
 \theta=\frac{\eps}{4(nL+1)},\qquad
 \eta=\frac{\eps M}{8n},\qquad h=\frac{\eps}{4K_0}.
\]
Divide stage $l$ into $m_l=\lceil\tau_l/h\rceil$ equal steps, skipping zero
durations. Their total number $m$ is at most $L+1/h$, and every step has
length at most $h$. For each gradient coordinate at a step, draw $J$
independent assignments of all rows other than $e$, where
\[
 J=\left\lceil\frac{2M^2}{\eta^2}
                 \log\frac{8nkm}{\eps}\right\rceil.
\]
Average the coupled marginal samples
$Y=f(R_{x,-e}+i\ind_e)-f(R_{x,-e})$, and set $w_{e,i}=\widehat w_{e,i}+\eta$.
Apply Lemma~\ref{lem:sampling-oracle} conditionally on the entire previous
history, with $\alpha=\eta$ and $\delta=\eps/(4m)$. A conditional union
bound over the $m$ steps shows that all upper-gradient inequalities hold
simultaneously with probability at least $1-\eps/4$.

Every direction has row masses $\ind_B$. Hence the entire trajectory is
feasible in $D$, and $q(t)$ belongs to $t$ times the matroid independent-set
polytope. On the simultaneous event, sum Lemma~\ref{lem:finite-step} over
all steps. At a stage boundary, Lemma~\ref{lem:smoothing} and
Lemma~\ref{lem:potential-estimates} bound the potential loss by $nM\theta$.
Initially all optimal thresholds equal $c(0,0)$, so
\[
 \Psi(0,r(0))=c(0,0)\OPT+(1-c(0,0))f(\boldsymbol0)
 \ge(\Gamma-\theta)\OPT.
\]
At the terminal time all thresholds vanish and $\Psi(x(T),0)=F(x(T))$.
Since the step lengths sum to $T<1$, their squares sum to at most $h$.
Using $M\le\OPT$ therefore gives
\[
 F(x(T))\ge\Gamma\OPT-\theta(1+nL)\OPT-2\eta n-K_0Mh
 \ge(\Gamma-3\eps/4)\OPT.
\]
Outside the event $F$ remains nonnegative. If $\Gamma\le\eps$, the claimed
expectation bound is immediate; otherwise the last lower bound is positive,
and $\Gamma\le1$ yields
\[
 \E[F(x(T))]\ge(1-\eps/4)(\Gamma-3\eps/4)\OPT
 \ge(\Gamma-\eps)\OPT.
\]
The step count, sample count, simplex projection, and greedy base selection
are polynomial in $n,k,1/\eps$ for the fixed certificate.

\subsection{Rounding without loss}

Lemma~\ref{lem:matroid-rounding} provides value-preserving rounding
for a single matroid without value queries. Here the retained bases give an explicit
implementation. First choose labels $I_e$ independently with probabilities
$x_{e,i}/q_e$ for $q_e>0$, and arbitrary labels for zero rows. For fixed $I$,
let $f_I(S)$ be the assignment value obtained by using label $I_e$ on
$e\in S$. This is a nonnegative submodular set function. Its ordinary
multilinear extension $F_I$ satisfies
\[
 \E_I[F_I(q)]=F(x).
\]
This identity follows row by row from affinity of $F$.

If the matroid rank is $r$, add $r$ free dummy elements and truncate the
direct sum to rank $r$. The saved expression
$q=\sum_{\text{steps}}\Delta\ind_B$ has total coefficient $T$; add the
all-dummy base with coefficient $1-T$. This is an explicit base
decomposition, and extend $f_I$ to ignore dummy elements.

For completeness, exchange rounding merges two bases $A,B$ with
coefficients $\alpha,\beta>0$. For $e\in A\setminus B$, symmetric base
exchange supplies $g\in B\setminus A$ such that both exchanged sets are
bases. With probability $\beta/(\alpha+\beta)$ replace $A$ by $A-e+g$;
otherwise replace $B$ by $B-g+e$. The represented point moves to
$z+\alpha(\ind_g-\ind_e)$ or $z-\beta(\ind_g-\ind_e)$, with mean $z$.
The ordinary multilinear extension of a submodular function is convex along
this exchange direction, since the second derivative is
$-2\partial_e\partial_gF_I\ge0$. Thus conditional expectation does not
decrease. Each exchange reduces the bases' symmetric difference; once they
agree, merge their coefficients. Repeated merging leaves one random base.
The required exchange can be found by enumerating elements and querying
independence. There are polynomially many exchanges, and no value queries.
Deleting dummy elements gives a feasible support $S$, with
\[
 \E[f_I(S)\mid I,x]\ge F_I(q),\qquad
 \E[f(\boldsymbol s)\mid x]\ge F(x).
\]
Together with the fractional bound this proves
Theorem~\ref{thm:certificate}.

% Recovered from the bibliography in the author-supplied English PDF.

\end{document}